\documentclass[runningheads]{llncs}
\usepackage[T1]{fontenc}
\usepackage{graphicx}
\begin{document}
\title{Critical Distribution System\thanks{supported by the CRISDIS project of the Czech Ministery of the Interior no. VI04000107.}}
%
%
\author{Anetta Jedlickova\inst{1}\orcidID{0000-0003-1239-4046} \and
Martin Loebl\inst{2}\orcidID{0000-0001-7968-0376} \and
David Sychrovsky\inst{2}\orcidID{0000-0002-4826-1096}}
\authorrunning{A. Jedlickova et al.}
%
\institute{Charles University, Prague, Czech Republic
\email{Anetta.Jedlickova@fhs.cuni.cz} \and
Charles University, Prague, Czech Republic
\email{\{loebl,sychrovsky\}@kam.mff.cuni.cz}\\
\url{https://kam.mff.cuni.cz/$\sim$loebl/}}
\maketitle              
\begin{abstract}
Distribution crises are manifested by a great discrepancy between the demand and the supply of a critically important good, for a period of time.
In this paper, we suggest a hybrid market mechanism for minimising the negative consequences of sudden distribution crises. 

\end{abstract}
\section{Introduction}
\label{s.int}

Distribution crises are extreme situations where the supply of a critically important good is so small that the actors realistically do not even have the necessary minimum available to ensure function. Such situations have two basic characteristics from a distribution point of view:

{\bf 1.} The {\em ethical fairness} of the distribution of supply of critical goods is important. According to the rules of fairness, which are declared in advance, it is possible to hypothetically distribute the supply of a critical commodity fairly among organizations.

In other words, let's imagine that buyers have {\em rights to purchase a fair amount} of the commodity in question. The amount of rights to purchase the commodity  corresponds to the amount of the commodity available on the market, and therefore the typically assigned rights are very few, i.e., significantly less than the minimum necessary to run the organizations.

Studying the determination of a fair allocation of scarce resources for each crisis individually according to the needs of individual facilities (sometimes e.g. facilities for seniors, sometimes paediatric facilities) in such a way as to maximise the benefit of the available material and to distribute its limited amount among all facilities according to their needs on the basis of a fair distribution model (i.e. a fair distribution of rights in a specific situation) is one of the important outcomes of our project.
In the next text, we assume that such a fair allocation of rights is already defined.

{\bf 2.} The existence of a {\em market for a critical commodity} is another essential aspect of the distribution. Lack leads to a significant increase in price, organizations try to obtain more than a fair amount to cover their needs, and this is always at the expense of other participants. A free market leads to a narrowing of distribution.
Experience shows that even centrally controlled distribution is not socially advantageous, for instance distributors do not have sufficient advantage in such an environment and try to leave it.
Since neither the free market nor the decision-making of the central authority is the most socially advantageous solution, we propose a hybrid model based on the autonomous behavior of the participants.
In this paper, we will partition the buyers into two main categories: {\em Active buyers} are market participants, who can be richer or more assertive or risk-seeking.
 {\em Passive buyers} are hesitant or risk-averse or poorer market participants. For simplicity, we will continue in this paper to use only the terms {\em active and passive buyers}, without specifying their motivation in more detail.
 
 \medskip

{\em Since the flow of money and critical goods in a free market during
a distribution crisis is mainly concentrated between sellers and active buyers, our aim is to extend the flow to passive buyers, in order to avoid situations where the distribution of scarce strategic material reaches only the most active or the wealthiest.}

\medskip

Our strategy may be thought of as analogous to the basic strategy of \cite{DKA} which, however, studies a different setting: maximising the potential of the marketplace itself to serve as a re-distributive tool.

\subsection{Main idea}
\label{sub.mi}

We will now describe the main idea of our solution. 
Our proposed system of distribution of scarce strategic goods  in a crisis is formed by a sequence of markets over time. In each market, all buyers have rights to purchase the commodity in question, the amount of rights of each participant corresponds to a fair distribution of the current offer.
In each market of the sequence, both the critical commodity and the rights are traded. There is only one limitation: at the end of each market, each buyer must have at least as many rights as the critical goods. 
There are three basic types of buyers. (1) The buyer buys exactly as much of the good on the market as it has assigned rights. (2) The buyer (active) buys more goods than it has assigned rights, and it has to buy additional rights from other market participants. (3) The buyer (passive) buys less goods than it has assigned rights and sells its remaining rights.


\medskip

{\em In this way, active buyers get more of the commodity, much as they would get on the free market.
But at the same time, they will pay for the rights to passive buyers, thereby expanding the flow of money in the market, and passive buyers will effectively get extra funds to buy the commodity or rights in the next market following the current one.
In each market, buyers will get new rights to buy the commodities on a fair basis and independently of the market that has just ended. The resulting benefit for all market participants, i.e. participating buyers and suppliers, increases by the widening of the flow of money and critical goods and is higher than by trading on the open market.}



\medskip

{\bf An overview of our distribution system:}
Our model consists of a sequence of markets over time: the first market takes place at time one, when it ends, the second market begins and so on. Each market has two parts:

1. Sellers announce the supply for the good, buyers announce the demand for the good, which greatly exceeds the announced supply.
On the basis of this input, the system assigns, according to pre-agreed rules of fairness, to each buyer a fair amount of {\em right to the good}.

2. Then both the good  and the right are traded with the unique condition: at the end of each market, each buyer must have sufficient amount of the right for the acquired good. The idea is that some buyers want to buy more of the good right away, so according to the rules of the market, they buy from hesitant buyers an amount of rights assigned to them. With this money saved, even those hesitant can buy the good in the next market, because the rights are recalculated before each market, and thus renewed.

\subsection{Frustration and Willingness to Pay}
\label{sub.fru}

We need to introduce some measures indicating how successful the proposed hybrid mechanism is. We recall that our goal is to widen the flow of the critical good in the time of a distribution crisis. 
For a single market, we define the {\em frustration} of a buyer $b$ as the ratio $a/r\leq 1$ where $r$ is the amount of the rights assigned to $b$ and $a$ is the difference between $r$ and the amount of the good purchased by $b$, if this is at least zero, and it is defined to be zero otherwise.
We note that the notion of frustration extends to normal times. In normal times, the amount of assigned rights is equal to the demand and moreover $b$ can buy all the goods it demands. Hence, its frustration is zero. Our concept of {\em frustration} is somewhat similar to  the concept of {\em deprivation cost} introduced in \cite{HPJ}.

Our aim is to show that in a sequence of markets of our system, {\em the frustration of each buyer becomes smaller than it would be in the free market (with analogously defined frustration)}. In order for that to happen it is needed that the frustrated buyers increase their {\em willingness to pay} (formally introduced in section \ref{s.market}).  Our proposed mechanism changes the willingness to pay by trading the assigned rights.

 \medskip

\subsection{Main Contribution} We suggest a new market mechanism for distribution crises based on autonomous behavior of participants. We initiate the study of properties of our system, in the middle of a crisis. A study of the beginning and the end of a crisis is postponed to future work. Next two chapters describe the model of fairness and the definition of the market. In the fourth section we show that a simple auction-based mechanism approximates the equilibrium of the Market in a strongly polynomial number of steps. In the fifth section we observe how the frustration decreases in the sequence of Markets, confirming the goal of the paper. Finally, in the last section we present some initial results on another (sellers-driven) market mechanism.

\section{Fairness}
\label{s.fair}

In this section we study the rules of the {\em ethical fairness} of the distribution of supply of critical goods. These rules need to be declared in the beginning of each crisis but may differ for different crises.
According to the declared fairness rules, it has to be possible to {\em hypothetically} distribute the available supply of a critical commodity fairly among participants. 
Clearly, a fair distribution depends on the total available supply and the individual needs of the participants. We will view a fair distribution as an algorithm. Its input consists of (1) the total available supply and (2)  the individual needs of the participants.
We need to address the issue of the {\em truthfulness of  the inputs of the fair assignment} since there are situations when it may be advantageous for a buyer to declare a higher than actual need or for a seller to declare lower than actual availability of the good. We are not aware of a general mechanism ensuring truthfulness in our setting. The issue of truthful reporting is discussed in subsection \ref{r1}.
Next we introduce some basic properties of a fair distribution. 

\subsection{Fair Distribution:  Basic Notions}

We denote the (finite) set of buyers by $B$. We will assume, for simplicity, that there is only one good to be distributed. We will denote by $G$ the set of available good; we denote by $|G|$ the size of $G$. If the good is indivisible then $G$ is a finite set. 
Each buyer $b\in B$ has its {\em truthful demand} $d_b$ for the good. A crisis is characterized by $d= \sum_{b\in B} d_b$ being much bigger than the size of  $G$. 


A {\em distribution} is a function $f= f(|G|,(d_b; b\in B))$ which, given the input consisting of $G$ and $(d_b; b\in B)$ determines for each $b\in B$, the amount $f(b)$ of $G$ assigned to $b$; clearly, $\sum_{b\in B}f(b)$ equals the size of $G$.
We say that $f$ is 

\begin{itemize}
\item
{\em Monotone.} If $b\in B$ and $x\leq x'$ then  $f(x,(d_b; b\in B))\leq f(x',(d_b; b\in B))$.
\item
{\em Consistent.} If $b,b'\in B$ then the distribution $f(f(b)+ f(b'), d_b, d_{b'})$ is $f(b), f(b')$.
\item
{\em Self-dual.} Loss, i.e., $d_b- f(b)$ is distributed in the same way as $f(b)$.
\item
{\em Fair} if it is monotone, consistent and self-dual.
\end{itemize}

Studying fairness differs for the divisible and indivisible goods. In this paper we restrict ourselves to the indivisible goods, but we have in mind the case of critical good needed in large quantities, like respirators, and thus it is almost liquid. Hence, for fairness concepts, we take inspiration from the divisible case.

\subsubsection {Case of Divisible Goods}
\label{sub.div}

We consider two distributions which satisfy the three properties above and are also efficiently computable: the {\em proportional distribution} (PD) and the {\em contested garmet distribution} (CGD).

\begin{definition}
\label{def.cg}
The constrained equal distribution for $(|G|, d_1, \ldots, d_n)$ is given by the following recursive procedure: if there is $i$ such that $d_i< |G|/n$ then assign $d_i$ to $i$ and distribute the rest to the remaining participants in the same way, else give $|G|/n$ to each participant. 

The {\em contested garmet} (CGD) is a distribution $f$ of a {\em divisible good}  which is monotone, consistent, self-dual and satisfies: If $|G|\leq (d_1+\ldots d_n)/2$ then $f$ coincides with the constrained equal distribution for $(|G|, d_1/2, \ldots, d_n/2)$.
\end{definition}

\begin{theorem} [Aumann, Maschler  \cite{AM}]
\label{thm.1}
 The CGD is unique and there is a polynomial algorithm that finds it.
\end{theorem}

\subsubsection{Optimisation of fairness}
\label{sub.oopt}

There are many fair distributions of divisible goods. We can choose a preferred one by {\em optimising} a function over the set of fair distributions.   A crisis can have many different forms and according to features of a crisis, we can adopt the fairness distribution by optimising different functions. 

From the ethical point of view, an approach maximising social welfare is preferred for a critical distribution, where we consider other
parameters of the current crisis, not only the need itself. In addition, for similar actors with similar needs, a uniform
approach is preferred. 
From a welfare perspective, an egalitarian welfare distribution does not require to distribute scarce resources equally, since equality of welfare is different from equality of resources and is morally valuable as a goal. 
Hence, our approach is consistent with both, the egalitarian ethical perspectives via equality of welfare in society, as well as with the utilitarian ethical perspectives that emphasize maximising the benefits produced by scarce resources, i.e., the greatest total welfare to members of society.


\begin{itemize}
\item
An example of such approach to fair distribution is first to structure buyers according to social preference for the particular crisis and then apply CGD in each part.
\item
Another possible approach looks almost like cheating: respect fairness properties and ethical preferences for the current crisis, and use an optimisation of fairness as a mechanism helping the social welfare, e.g., minimise the total (maximal individual respectively) frustration. For example, assign less rights for the good to buyers with highest willingness to pay. 
\end{itemize}

The features of the adapted fair distribution should be generally accepted in the society and the second approach is questionable from this point of view. Optimising fairness of allocating resources is extensively studied nowadays, see e.g. \cite{S}, \cite{V}.
For a survey of fair allocation of scarce resources, see \cite{T}.

\subsubsection{Adaptation to the distribution of indivisible good}
\label{sub.indist}

Distributions of divisible goods can be naturally adopted to distribution of indivisible good for instance as follows: 
(1) Find the {\em fractional} distribution as if the good is divisible,
(2) Round down the fractional distribution for the participants,
(3) Distribute the surplus reflecting the society preferences by a chosen optimisation mechanism. A simple possibility is to linearly order the participants and round up the fractional solution for the participants in an initial segment.

As pointed out by Ron Holzman, these adaptations do not preserve the good properties of the original method in the divisible setting but may help the central authority to {\em enhance (its understanding of) the fairness} by treating participants in an unequal but socially beneficial and accepted way.

\section{A Sequence of Markets for one scarce indivisible good}
\label{s.market}

In the previous section we studied mechanisms for a {\em fair distribution} of the limited available amount of the good among all participants, i.e., a fair distribution of {\em rights to the good}.

In this section, each critical episode will consist of a finite sequence of {\em Markets} which happen one after another in time. The participants are divided into buyers and sellers depending on whether they produce or consume the good, not the right. We will assume that for each Market of the sequence, a fair allocation of rights to the buyers is already defined. The good and the right are traded with the only condition that after each market of the sequence ends, each buyer needs to have at least as many items of the right as the items of the good. It remains to specify Markets in detail.

\subsection{A view of a single Market}
\label{sub.m}
Let us describe how a market with items of one scarce commodity called the Good and with items of the Right could work. 

The first step towards a practical implementation of a functioning crisis distribution system based on our suggested mechanism is to create a Certified Public Portal (CEP) facilitating trading both in normal and critical times. 
For example by the Czech law, in various legal crisis settings the authorities in charge can require all dealings concerning strategic goods in a prescribed portal, and all the production and storage reported.

 The Certified Public Portal (CEP) sends out a message to central authorities that a distribution crisis is happening for a commodity called Good. 

A central authority reacts by declaring a crisis and declares that CEP is a designed portal for the distribution of the Good during this crisis.
The central authority declares the schedule for trading in the portal:

a. Time period for declaring the demand and the offer in the portal in which trading is frosen.  Mechanisms for truthful declarations need to be in place. 

b. The portal freezes demands and offers and issues the items of the Right according to the agreed on fair distribution.

c. The portal opens trading with the Good and the Right. 

d. End of trading: All the buyers who have more items of the Good than items of the Right are penalised.  Unmatched items of the Right disappear. A schedule of next trading is declared.

If new supply of items of the Good appears during the trading, CEP flexibly recalculates and distributes new items of the Right to the buyers. 

There needs to be a protocol for maintaining {\em money} obtained by selling items of the Right. It is an ethical requirement that such money stays in the system and is used for buying items of the Good or the Right in the future Markets. This rule also supports enhancing individual willingness to pay by the mechanism of trading the rights. We introduce the third commodity called the Money in the formal definition of Market below. 

\subsection{Definition of a Single Market of a Critical Episode}
\label{sub.m}
\begin{itemize}
\item
 The participants of each Market of the sequence of Markets of a critical episode are the same, and are partitioned into {\em buyers} and {\em sellers}. We denote by $B$ the set of the buyers and by $S$ the set of the sellers. 
 \item
 In each Market of the sequence, three indivisible commodities will be traded: {\em Good, Right, Money}. We denote by $G$ the set of all the items of the Good, by $R$ the set of all the items of the Right and by $M$ the set of all the items of the Money.  The trading is facilitated by selling and buying items of commodities. In these acts, the price of the same commodity may differ. However,
 \item
 We require that the price of each unit of Money, i.e., of each element of $M$, is always constant. 
 We denote by $\delta$ the price of the unit of the Money and for simplicity we assume $\delta = 1$.  
 
 Another reason to have Money as a commodity is that the utility for Money differs among the participants and we will use this fact for studying the considered sequence of markets of a critical episode. For similar treatment of Money, see \cite {DKA}.

\item
We denote by $u_p(H,x)$ the utility of  $x$ items of set $H$ for a participant $p$. All $u_p(H,x)$ in this paper will be monotone.

\item
We will further assume that for a buyer $b$, $u_b(G,x)+u_b(G,y)\geq  u_b(G,x+y)$ which is a natural assumption for a critical commodity. 

\item
We will further have that for each participant $p$, $u_p(M,x)$ depends linearly on $x$.

\item
 Sellers have a positive utility only for Money.
 
 \item
The dynamics of the Market is  determined by $u_b(M,x)$ and $u_b(G,x)$, $b$ buyer. These individual utilities are crucial for studying the Market and its outcome and they are not disclosed. 
\item
The {\em Willingness to pay} of a buyer $b$ is an important function which, unlike the utility functions, can be observed in the Market. The willingness to pay  associates, with a non-negative integer $x$, the number $w_b(x)$ of items of the Money of the same utility as $x$ items of the Good. Formally,  $w_b(x)= u_b^{-1}(M, u_b(G,x))$.
 \item
 In the Market, each participant has an initial endowment of the commodities, and trades them according to its utility function. Let us denote by $m^b, r^b$ the initial endowments of Money and Right for a buyer $b$.
  \item
 We assume that for each Market, initial endowments of Good, Right and Money, i.e., subsets of $G$, $R$ and $M$ for the buyers and the sellers are known in the market's beginning.  {\em Why can we assume this?} 

For the Good this follows from truthful reporting of availability of the Good by the sellers.

For the Right this follows from truthful reporting of the demand by the buyers which determines the distribution of the items of the Right to each buyer.

For the Money, it is practical to assume that each of the participants has initially sufficiently large amount of items of Money. This allows us to model broader situations. 

\item
For each buyer, the relation of its utility for Good and for Money is crucial. We will assume:
\begin{enumerate}
\item
$m^b\geq 4r^b$,
\item
For each $x\leq r^b$, $u_b(G,x)>  2u_b(M,x)$
\item
For each $x\geq 1/2m^b$, $u_b(M,m^b)> u_b(M,m^b-x)+  u_b(G, x)$.
\end{enumerate} 
\item
In order to study the behavior of a Market, we introduce the notion of a {\em solution} and an {\em equilibrium}. A {\em solution} consists of (1) a price vector giving the price per item for each of the three commodities, and (2) a partition of a subset of the union of all the initial endowments into {\em baskets}, one basket for each participant.

Given a solution, we say that the assigned basket of a participant is {\em feasible} if it satisfies: (1) the total price of the basket is at most the total price of its initial endowment and (2) in the basket, the number of items of Good is at most the number of items of Right. A solution is {\em feasible} if all the baskets are feasible.

A solution is an {\em equilibrium} if for each participant, (1) the total price of its basket is equal to the total price of its initial endowment and (2) its utility of  its basket is maximum among all feasible sets of items.

\end{itemize}

\subsection{Strategy and coalition proofness}
\label{r1}
 
Our strategical aim is that the rights are sold as soon as possible so that passive buyers get funds without delay and can soon buy the Good in subsequent markets. 
We require that after each Market of the sequence, for each buyer there is a check whether it has at least as many items of the Good as of the Right. 
Is this a feasible process?  We list features of the proposed system supporting the strategy and coalition proofness.
(1)  The distribution system is implemented in a certified portal, the history of trading provides indirect mechanisms for (checking) truthful reporting of demand and supply,
(2) The complete history of trading of the Right is known by using tailored data structures,
(3) We suggest to have some special participants which help ensuring strategy and coalition proofness, e.g., a participant in charge of the rights.
(4) It is possible to implement the Market using anonymous mechanisms. In this paper, we study one such {\em buyers-driven} mechanism, see Definition \ref{def.auc} and Corollary \ref{thm.auc}. Studying {\em sellers-driven} mechanism is initiated in the last section.

\begin{definition}
\label{def.auc}
The {\em Couple mechanism} is the following way to implement the Market which is driven by the activity of the buyers: (1) In the beginning, the central authority assigns the items of Good to buyers arbitrarily so that each buyer has the same number of items of Good and Right. Each buyer arbitrarily pairs the items of Good and Right into items of {\em Couple}. (2) The buyers subsequently trade the items of Couple. (3) In the end of trading the prices are cleared, i.e., the price of each item of Couple is equally divided between Good and Right and the sellers obtain the resulting money for their items of Good, and the buyers obtain the resulting money for their items of Right; 
(4) An {\em equilibrium} for this mechanism is defined as the market clearing for the trading part of the mechanism.
\end{definition}

\section{Computing Market equilibrium}
\label{s.am}
In this section we gather information on the defined Market in particular we introduce an efficient algorithm for finding an approximate equilibrium.

\subsection{An Auction-based Algorithm}
\label{r5}
We describe an efficient auction based algorithm to find an approximate equilibrium for the Market. The approximation ratio of the algorithm is $1- 2\epsilon$ where $\epsilon$ satisfies, for each buyer $b$,
 $1> \epsilon > 2/m^b$.

For the purpose of the algorithm, we introduce one more commodity of indivisible items called the Couple. 
Each item of Couple is a pair $(s,t)$ where $s$ is an item of Good and $t$ is an item of Right. We denote by $u_p(C,x)$ the utility of  $x$ items of Couple for a participant $p$, and we let $u_p(C,x)= u_p(G,x)$.
The algorithm {\em auctions}  items of Couple. We will denote the current price of one item of Good (Right, Money, Couple respectively) by $\pi_G$ ($\pi_R$, $\pi_M$, $\pi_C$ respectively).

\medskip\noindent
{\bf The algorithm description.}  
\begin{itemize}
\item
Initially, we let $\pi_M= \pi_G= \pi_R:= 1, \pi_C:= 2$ and each buyer gets the surplus money covering its initial endowment. 
\item
The algorithm is divided into {\em iterations}.  During each iteration, some items of Couple are sold for $\pi_C$ and some for $(1+\epsilon)\pi_C$, and analogously for Good and Right. The price $\pi_M$ is always constant equal to $1$. 
\item
Each iteration is divided into {\em rounds}. An iteration ends when the price of Couple is raised from $\pi_C$ to  $(1+\epsilon)\pi_C$. 
\item
Round: consider buyers one by one. Let buyer $b$ be considered. Let us denote by $o^b$ the number of items of Couple $b$ currently has, and by $o_+^b$ the number of items of Couple $b$ currently has of price $(1+\epsilon)\pi_C$. 
\item
Let $S^b$ be $b$'th optimal basket given the current price $\pi_C$, i.e., a set of items of Couple and of Money of max total utility which $b$ can buy with its cash plus $\pi_Co^b$. Let $s^b$ denote the number of items of Couple in $S^b$.

\begin{enumerate}
\item
If $s^b< o^b$ then $b$ does nothing, the algorithm moves to the next buyer. 

[if this happens then the current basket of $b$ is optimal for the previous price $\pi_C/ (1+\epsilon)$ and $o_+^b= 0$.]
\item
If $s^b\geq  o^b$ then $b$ buys items of Couple via the {\em Outbid}. 
\end{enumerate}
 
\medskip\noindent
{\bf OUTBID:}

\begin{itemize}
\item
The system buys as many as possible but at most $s^b-  o_+^b$ items of Couple for price $\pi_C$ and sells them to $b$ for price $(1+\epsilon)\pi_C$. First, it buys from $b$ itself. 
\item
An alternative for buying the items of Couple is to buy separately items of Good and Right and compose them into items of Couple.  This happens when some units of Right and (necessarily the same amount of units of) Good are not yet coupled in previous tradings. We observe that this happens only if they are available for the initial price from the participants. In this situation, the system again buys items of Right first from the buyer $b$.
\item
However, the system pays nothing if it buys items of Right from an initial endowment of a buyer for the initial price since the payment is already in the surplus money.
\end{itemize}

\item
If no more Couple is available at price $\pi_C$  after the Outbid then the current round and iteration terminate, $\pi_G:= (1+\epsilon)\pi_G$ , $\pi_R:= (1+\epsilon)\pi_R$ , 
$\pi_C:= (1+\epsilon)\pi_C$ and the surplus money are updated:  
everybody who had Good or Right in its initial endowment gets extra surplus money, $\epsilon\pi_G$ or $\epsilon\pi_R$.
\item
If a round went through all players, the algorithm proceeds with next round. 
\item
When nobody wants to buy new items of Couple, the whole trading ends. The system takes all items of Money from the buyers, sells them to the buyers and sellers and keeps whatever items remain.
\item
The OUTPUT consists of (1) the collection of the final baskets of each participant and (2) the price-vector of the terminal prices  $\pi_G, \pi_R, \pi_C$.
\end{itemize}
 
\medskip
 
 \subsection{Analysis of the Algorithm}
 \label{sub.anall}
 
\begin{itemize}
\item
We recall that we denote by $m^b, r^b$ the initial endowments of Money and Right for $b\in B$, and we assume: 

(1) $m^b\geq 4r^b$,

(2) for each $x\leq r^b$, $u_b(C,x)>  2u_b(M,x)$ and

(3) for each $x\geq m^b/2$, $u_b(M,m^b)> u_b(M,m^b-x)+  u_b(C, x)$.
\item
We denote by $m, r, g$ the total initial endowments of Money, Right and Good and recall that $r= g$.
 \item
 We note that at each stage of the algorithm, the total surplus is at most $2m$: it is true in the beginning by assumption (1), the surplus is gradually decreased during each iteration and at the end of each iteration,  deleted funds are given back.
 \end{itemize}
 
 {\em Claim 1.}
 In the first iteration, all items of Good and Right are paired.
 \begin{proof}
 By assumptions (2) all the buyers prefer to buy at least the fair amount of items of Couple for the initial price; by assumption (1) there is enough cash in the initial surplus of each buyer to do it.
\end{proof}
 
  {\em Claim 2.}
 After the end of the first iteration: (1) a buyer owes to the system only money for items of Money in its initial endowment and (2) total cash among participants  is always at most $m$.
 \begin{proof}
 The first part follows from Claim 1 since all items of Good and Right are sold and bought at the end of the first interaction. For the second part we note that among sellers, the total cash is $g\pi_G$ since all items of Good were sold in the first iteration and among buyers, the total cash is at most $m-g\pi_G$ since the buyers payed for the items of Good and there is no cash left from the initial endowments of Right since all items of Right were sold and bought in the first iteration.
 \end{proof}

{\em Claim 3.}
  The number of rounds in an iteration is at most $2+ |B|$.
 \begin{proof}
 We observe that in each fully completed round, either none of the buyers buys items of Couple and the trading ends, or none of the buyers buys items of Couple in the next round and the trading ends or at least one buyer acts for the last time in this iteration: if in the current round every buyer buys items of Couple only from itself then in the next round nobody buys since nobody got additional cash. Hence let a buyer $b$ buy items of Couple from another buyer in the current round. It means that $b$ gets no additional cash in this iterations since it has no items of Couple for $\pi_C$, otherwise it would have to buy these first by the rules of the outbid and the current round is the last active round for $b$.
  \end{proof}
 
  {\em Claim 4.}
 The total number of iterations is at most $1+\log_{1+\epsilon}m$.
  \begin{proof}
 Each iteration raises the price of Couple by the factor of  $(1+\epsilon)$ and the max price per unit of Couple cannot be bigger than the total surplus.
 \end{proof}
 
 {\em Claim 5.}
 Relative to terminating prices, each buyer or seller gets a basket of utility at least $(1-2\epsilon)$ times the utility of its optimal feasible basket.
  \begin{proof}
  (1) Buyers owe nothing to the system since after the end of the trading they keep only the items of Money they can buy with their remaining cash.
 
 (2) After the end of trading and buying items of Money, each participant is left with less than $1$ dollar by the second part of Claim 2. 
 
(3) The basket of each seller is optimal since all items of Good were sold. 

(4) The only reason why the basket of buyer $b$ is not optimal is:
 For some items of Couple, $b$  payed $(1+\epsilon)\pi_C$ where $\pi_C$ is the terminal price of Couple. 
  Let $c$ denote the  total number of items of Couple in b's {\em optimal basket} and let $m_b$ denote the total number of items of Money in b's {\em optimal basket}. The utility of the {\em optimal basket} of $b$ is thus $u_b(C,c)+ u_b(M,m_b)$. By assumption (3), $m_b> c\pi_C$.
 
 In $b$'s {\em terminal basket},  there are $c$ items of Couple and at least $m_b- \epsilon\pi_Cc-1$ items of Money. First let $ \epsilon\pi_Cc\geq 1$.
The utility of $b$'s {\em terminal basket} is thus, using the assumption on the linearity of the utility of Money, at least 
$u_b(C,c)+ u_b(M,(1-2\epsilon)m_b)=  u_b(C,c)+ (1-2\epsilon) u_b(M,m_b)$. 

Secondly let $ \epsilon\pi_Cc< 1$.
The utility of $b$'s {\em terminal basket} is thus at least 
$u_b(C,c)+ u_b(M,(m_b-2))$ and Claim 5 holds since we assume $1> \epsilon > 2/m^b$.
 \end{proof}
 
 This finishes the analysis of the algorithm. We note that the outcome of the algorithm has an interesting feature: the terminal price of Right is equal to the terminal price of Good. The analysis is summarized in the next theorem.
 
 \begin{theorem}
 \label{thm.main}
 The following holds.
  \begin{enumerate}
 \item
 Assuming an act of a buyer in each round takes a constant time, the auction-based algorithm is strongly polynomial in the input size.
 \item
 For each participant, its basket assigned by the auction-based algorithm is feasible and its price plus $1$ is bigger than the total price of its initial endowment. 
 \item
 Relative to terminating prices, each buyer or seller gets a basket of utility at least $(1-2\epsilon)$ times the utility of its optimal feasible basket.
 \end{enumerate}
 \end{theorem}
 \begin{proof}
(1)  It follows from Claims 3, 4 that the time complexity of the auction-based algorithm behaves asymptotically as $|B|^2(1+\log_{1+\epsilon}m)$.
(2) follows from (2) of the proof of  Claim 5. (3) is Claim 5.
 \end{proof}

Claim 1 of the above analysis of the auction-based algorithm means that the algorithm also provides an implementation of the Couple mechanism of Definition \ref{def.auc}.
Hence, we get the following

\begin{corollary}
\label{thm.auc}
There is an auction-based implementation of the Couple mechanism which leads in a strongly polynomial (in the input size) number of steps to an approximate equilibrium.
\end{corollary}

\section{Development of Willingness to Pay and Frustration in the Sequence of Markets}
\label{s.wp}
The Markets happen subsequently in a sequence. How the parameters of subsequent Markets of the sequence change? We distinguish two regimes, (1) Deep in the crisis and (2) At the beginning or approaching the end of the crisis.

In this paper we study only the regime (1).
This enables us to assume that the individual utility of the Good does not change in subsequent Markets. 

What changes is the individual utility of the Money: buyers who sell items of the Right enter the subsequent Market with more Money and the new items of Money can be used, by the rules, only to buy items of Good or Right. Hence, the utility function of Money changes for these buyers.
We note that, since the individual utility of the Good is unchanged, the change of the individual utility of the Money can equivalently be described as a change of the willingness to pay.

\begin{definition}[development of the willingness to pay]
\label{def.ww}
Let buyer $b$ sell, in the current Market, items of the Right for the total price $y_b$. If $b$ buys Right then we let $y_b= 0$. Let $w_b$ denote its willingness to pay in the current Market. Then for the next Market, its {\em potential willingness to pay}, denoted by $w'_b$, is $w'_b(z):= w_b(z)+ y_b$, $z$ arbitrary. 
\end{definition}

\begin{theorem}
\label{thm.wp}
Let us consider a sequence of Markets satisfying (1) the total supply, the individual demand and the individual utility of Good does not change in the sequence, (2) the Markets are implemented by the auction-based algorithm and (3) after a Market of the sequence ends, the willingness to pay in the subsequent Market is equal to the potential willingness to pay. Then in all but possibly the first Market, each individual frustration is at most $1/2$. 
\end{theorem}
\begin{proof}
Let a Market $M(i), i\geq 1$ of the sequence ended and let us consider the next Market $M(i+1)$. By the assumptions of the theorem, the auction-based algorithm repeats the steps of Market $M(i)$. After the final step of the auction for $M(i)$,
the willingness to pay of the buyers with zero frustration in $M(i)$ (let us call them {\em happy}) is saturated.

However, the buyers with positive frustration in $M(i)$ (let us call them {\em frustrated}) remain active since they acquired additional funds in $M(i)$. Let $b$ be such a frustrated buyer. We recall that its willingness to pay, given the current price of the Couple (which is equal to the final price of the Couple in $M(i)$), covers the number of items of the Couple which is equal to the number $r_b$ of items of the Right $b$ currently sold (which is equal to the number of items of the Right $b$ sold in $M(i)$). Let us denote by $S$ the set of $r_b$ items of the Couple which contain the items of the Right buyer $b$ sold so far in $M(i+1)$. 

Buyer $b$ buys the Couple of $S$ at an increased price which in turn frees funds of active buyers who may buy back. During this process the frustration of $b$ can only go down. Hence, we only need to consider the case that the frustration of $b$ is strictly bigger than $1/2$. Let $0< r'_b< r_b$ be such that $r_b-r'_b$ is equal to the half of the number of assigned rights to $b$. 

Let us assume that $b$ doubled the price of the Couple in its first bid, and buys $r'_b$ additional items of the Couple of $S$. Buyer $b$ only needs to buy items of the Good by which $b$ spends all $y_b$ additional items of the Money it got from the previous Market $M(i)$: $2r'_by_b/r_b$ items of the Money for buying $r'_b$ items of the Good from $S$, and $(r_b-2r'_b)y_b/r_b$ items of the Money which are needed to increase the price of items of the Good $b$ already has.

We observe that happy buyers from whom $b$ bought new items of the Couple are not willing to buy back since: 

They obtain $2r'_by_b/r_b$ items of the Money for the sold $r'_b$ items of the Couple of $S$ and need $2(r_b-r'_b)y_b/r_b$ items of the Money to increase the price of their remaining $(r_b-r'_b)$ items of the Couple in $S$; clearly by the definition of $r'_b$, $(r_b-r'_b)\geq r'_b$. 

Summarizing, the frustration of $b$ is at most $1/2$ by spending in addition only the Money obtained from selling the Right in $M(i)$.

\end{proof}

\section{Seller-Driven Market: Preliminary Results}
Next, we numerically study more realistic seller-driven Market mechanism which is a variant of the double-auction with two kinds of auctioned goods. A continuation of this study is a work in progress \cite{CFLS}.

A sequence of the below described Markets can be thought of as a multi-agent game. We use a state of the art deep reinforcement learning algorithm to approximate optimal strategies of the players. The algorithm tries to find a strategy for each participant which maximises the expected future sum of its utility. 
Each agent's strategy is parameterized by a deep neural network. The strategy (policy) is trained by Twin Delayed Deep Deterministic Policy Gradient (TD3) \cite{TD3}.
Let us begin by defining the the Market mechanism.

\begin{definition}[Sellers Market Mechanism]
\label{def.market_mech}
In the beginning of each Market of the sequence,
(1) each buyer $b$ receives $m_b$ additional units of Money, and each seller receives $g$ additional units of Good.
(2) Each seller declares the amount of Good for sale along with the selling price.
(3) Each buyer $b$ declares the demand $d_b$ for Good and (4) the resulting items of Right are distributed according to the CGD, see Definition \ref{def.cg}.
Then, (5) each buyer declares the amount of Rights he is willing to sell along with the selling price.

Each buyer, given the declared quantities and prices, orders a number of offered items of the Good and the Right by declaring its desired prices and volumes for the Good and the Right.
The Market randomly pairs the compatible offers and orders, starting with the cheapest offers. In order to buy items of Good, the buyer needs a sufficient amount of items of Right. Therefore, a buyer first buys items of Good and pairs them with its items of Right until it has no Right left. Then, it buys equal amount of Good and Right. Finally, each buyer $b$ consumes at most its demand $d_b$ of purchased Good and keeps the possible surplus for the future.
\end{definition}

\subsection{Utilities}\label{s.rl_u}

Next, we describe the utilities of participants in a sequence of Markets which models a period of a developed distribution crisis. For a better understanding, we measure the amount of Good and Money in {\em units} in this section, similarly as if they are divisible.

The utility of a seller $s$ {\em in the end of each Market of the sequence} consists of (1)  the amount of Money $s$ received during that Market (denoted by  $\Delta M_s$) and (2) a small negative utility (denoted by  $c_{\rm store}$) per unit of Good the seller still keeps which models the cost of storing the Good and motivates sellers to sell the Good promptly. 
At the end of the whole period, the sellers also get some small utility (denoted by  $c_{\rm end\ supply}$) per unit of Good they are left with since they would be able to sell the remaining Good eventually.

Utility of the buyers reflect their need to keep a steady supply of Good throughout the period. At the end of each Market, their utility is a linear function of the number of items of Good they have, clipped to some minimum if they have no Good, and maximum if their demand is saturated. At the end of the period, they also get some small utility for the Money they have. Formally, 

\begin{definition}[Utility]
Consider a sequence of Markets $X_1, \ldots, X_T$. For $t= 1, \ldots, T$ let (1) $d_b(t)$ denote the demand of buyer $b$ in market $X(t)$, (2) let $M_p(t)$ and $G_p(t)$ be the amount of Money and Good participant $p$ has at the end of Market $X_t$, (3) let $\Delta M_p(t)$ denote the amount of Money $p$ earned in $X(t)$, and (4) we also let $D(T)=1$ and  $D(t)=0$ for $t< T$. 

For a seller $s$,
\begin{equation}
    u_s(t) = \Delta M_s(t) + c_{\rm store} G_s(t) + D(t) c_{\rm end\ supply} G_s(t),
\end{equation}

For a buyer $b$, 
\begin{equation}
    u_b(t) = c_{\rm in\ stock} \min\left\{1, \frac{G_b(t)}{d_b(t)}\right\} + c_{\rm missing} \max\left\{0, 1 - \frac{G_b(t)}{d_b(t)}\right\} + D(t) c_{\rm money} M_b(t).
\end{equation}
\end{definition}

\subsection{Results}\label{s.rl_results}
In this section, we present results of a particular instance of a sequence of the above described Markets. We aim to model a situation in which there is a large discrepancy between the active and passive buyers. Specifically, we consider a sequence of $T = 10$ Markets with four buyers and four sellers. The demands and incomes of buyers do not change for $t = 1, \ldots, 10$ and are given in Table \ref{tab.buyer_stats}. The sellers receive $g = 1/4 + \mathcal{N}(0,1/40)$ units of Good per Market, where $\mathcal{N}$ is the normal distribution. At the start of the simulation, the participant have no Money or Goods.

\begin{table}[t]
\begin{center}
\begin{tabular}{ c|c|c|c|c } 
 \hline
 Buyer & 1 & 2 & 3 & 4 \\ 
 \hline \hline
 $m_b$ & $4/4$ & $5/4$ & $6/4$ & $1/4$ \\
 $d_b$ & $1/2$ & $1/2$ & $1/2$ & $5/2$ \\ 
 \hline
\end{tabular}
\end{center}
\caption{Demands and earnings of the buyers.}
\label{tab.buyer_stats}
\end{table}

We use the following values for the constants: $c_{\rm store}=-0.5$, $c_{\rm end\ supply}=0.1$,  $c_{\rm in\ stock} = 2$, $c_{\rm missing}=-1$ and $c_{\rm money}=0.1$. These ensure that the utility is mainly influenced by the primary motivation of each participant. The choice $c_{\rm in\ stock} = -2c_{\rm missing}$ aims to keep the mean utility near zero.

The first three buyers together have $93.75\%$ of the Money and thus in the free market will receive $93.75\%$ of the Goods. In contrast, the fair distribution following CGD is uniform if all Goods from the previous Markets are sold. If the sellers choose to offer more then the amount corresponding to uniform distribution, the CGD allocates the surplus Rights to the passive buyer first.

\begin{figure}[t]
    \centering
    \includegraphics[width=0.45\textwidth]{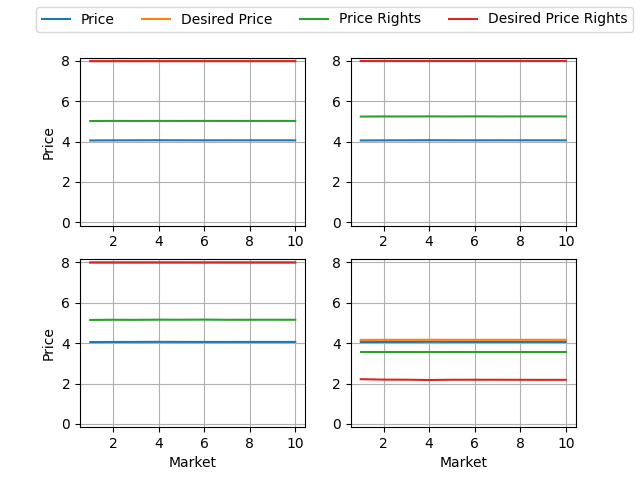}
    \includegraphics[width=0.45\textwidth]{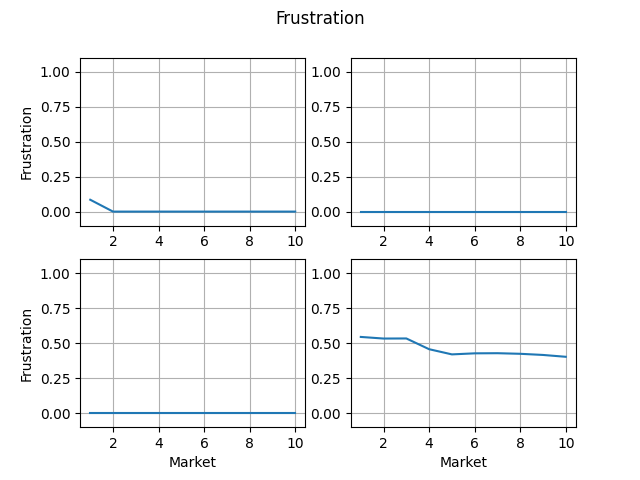}
     \label{fig:prices}
\end{figure}

To train the strategy, the agents played $36\ 000$ games, but the system stabilized after only $\approx 20\ 000$ games. The final prices for each Market are shown in Figure \ref{fig:prices}. Let us focus on the Goods first. The desired price for all buyers is larger than the selling price, thus the price set by sellers is acceptable for the buyers. The selling price is higher than the market clearing price\footnote{The market clearing price is four in this case because the sellers get one unit of Goods in total per step, while the buyers get four units of Money.}, which implies that some Goods do not get sold\footnote{Additionally, the sellers choose not to offer all Goods they have.}. Indeed, only $\approx 90\%$ of the Goods are sold. A possible explanation: during training, the agents are forced to explore and thus sometimes take sub-optimal actions. From the sellers' perspective, the selling price thus needs to be higher to ensure buyers are not left with extra Money.

The price of Right generally follows the price of Good and is larger for the active buyers. This suggests that Money the passive buyer receives for selling Right is more important for him compared to the other buyers and so he is willing to decrease the price in order to increase the chance a trade will be completed. As one would expect, the desired price of Right of the last buyer is lower than the selling prices, since he aims to sell his Rights. 

The active buyers buy on average $\approx 0.05$ units of Right per step, which is $\approx 20\%$ of the Right assigned to them by the uniform distribution. Specifically, the buyers bought $(0.17, 0.33, 0.7, 0)$ units of Right in total. Figure \ref{fig:prices} shows the evolution of frustration during the sequence of Markets. For the active buyers the frustration is zero, while for the passive buyer it decreases over time. At the end of the sequence, his frustration is $\approx 0.4$ in agreement with Theorem \ref{thm.wp}.

The amount of Goods each buyer purchased during the sequence of Markets was $(2.47, 2.73, 3.1, 1.46)$. The passive buyer was thus able to obtain $\approx 15\%$ of the total amount of Good, while in the free marker he would have Money for only $\approx 6\%$.

Figure \ref{fig:prices} shows, Left: selling and desired prices for each Market after convergence. The buyers are arranged from left to right and then from top to bottom.
   The selling price of Goods is averaged over all sellers. Right: Evolution of frustration of each buyer during the sequence.

\section{Conclusion} In this paper, we define and study a new market mechanism for distribution crises based on autonomous behavior of participants. The main ingredients of the mechanism are the model of fairness and the definition of  Market. We show that a simple auction-based Couple mechanism implementing the Market approximates the equilibrium of the Market in a strongly polynomial number of steps. We also confirm that the frustration decreases in the sequence of Markets implemented by the Couple mechanism, confirming the goal of the paper. 

Finally, in the last section we present some initial results on another (sellers-driven) market mechanism implementing the Market. More thorough study of this mechanism is work in progress.

\subsubsection{Acknowledgments} We regularly discuss with several people: Jan Bok, Martin \v{C}ern\'{y}, Jakub \v{C}ern\'{y}, Sameer Desai, Ji\v{r}\'{i} Fink, Iva Holmerov\'{a}, Ron Holzman, Zbyn\v{e}k Loebl, Jitka Soukupov\'{a}.

%
%
%
%

\end{document}